\documentclass[journal]{IEEEtran}
\usepackage{graphicx}
\usepackage{amsmath}
\usepackage{subfigure}
\usepackage{rotating}
\usepackage{multirow}
\usepackage{array}
\ifCLASSINFOpdf

\else

\fi

\begin{document}
\title{Modeling Routing Overhead Generated by \\ Wireless Proactive Routing Protocols}

\author{\IEEEauthorblockN{Nadeem Javaid, Ayesha Bibi, Akmal Javaid, Shahzad A. Malik\\\vspace{0.4cm}}
%    \IEEEauthorblockA{ \{nadeem.javaid,djouani@univ-paris12.fr\}}\\
        Department of Electrical Engineering, COMSATS Institute of\\
        Information Technology, 44000, Islamabad, Pakistan. \\
        nadeemjavaid@comsats.edu.pk\\

     }
\vspace{-2cm}

%
%% make the title area

\maketitle

\begin{abstract}
%\boldmath
In this paper, we present a detailed framework consisting of modeling of routing overhead generated by three widely used proactive routing protocols; Destination-Sequenced Distance Vector (DSDV), Fish-eye State Routing (FSR) and Optimized Link State Routing (OLSR). The questions like, how these protocols differ from each other on the basis of implementing different routing strategies, how neighbor estimation errors affect broadcast of route requests, how reduction of broadcast overhead achieves bandwidth, how to cope with the problem of mobility and density, etc, are attempted to respond. In all of the above mentioned situations, routing overhead and delay generated by the chosen protocols can exactly be calculated from our modeled equations. Finally, we analyze the performance of selected routing protocols using our proposed framework in NS-2 by considering different performance parameters; Route REQuest (RREQ) packet generation, End-to-End Delay (E2ED) and Normalized Routing Load (NRL) with respect to varying rates of mobility and density of nodes in the underlying wireless network.

\end{abstract}


\IEEEpeerreviewmaketitle

\vspace{-0.3cm}
\section{Introduction}

\IEEEPARstart{R}{outing} is a necessary but challenging goal in Wireless Multi-hop Networks (WMhNs). The dynamic nature of the wireless medium leads to frequent disconnection of links and then routes among different source-destination pairs. In many scenarios, such as disaster response, the network is supposed to perform routing to ensure the application that the delay caused is acceptable for information delivery.

Routing protocols are divided into two main categories; reactive and proactive on the bases of their routing behavior. Reactive protocols perform routing operation when request for a route is arrived, therefore, also known as on-demand routing protocols. While in proactive routing protocols, nodes continuously attempt to be aware of their neighbors and then whole topology.

Proactive routing protocols periodically exchange topological information, therefore, each node contains the whole network information in its route table. Along with this periodic activity, some protocols may or may not employ any of the necessary route (re)calculating operations, like, \textit{trigger updates ($RU\_Tri$)} and \textit{periodic updates ($RU\_ Per$)} for updating route status. The chosen proactive protocols are DSDV [1], FSR [2] and OLSR [3]. Reasons to select these proactive protocols include: DSDV is ideal for small no-of-nodes/no-of-data-flows during varying rates of mobility, FSR is ideal for very dense and dynamic network and OLSR is designed for static and dense networks [4]. In [5], we have investigated three widely used reactive protocols, so, in this work we focus proactive ones.

Proactive protocols face two major problems of increased delay and bandwidth consumption in WMhNs. There are two main periodic maintenance operations: i) individual link maintenance, and ii) overall route maintenance using beacon messages. These operations provide convergence, but their implementation leads to bandwidth issue. On the other hand, lack of these operations results low convergence, if alternative solution is either unavailable or inefficient.

	In wireless environment, to achieve higher throughputs, routing protocols implement different strategies. For example, in high dynamic situation, when there are frequent link breakages, DSDV sends $RU\_Tri$, if any link of an active route is broken. FSR, due to graded-frequency technique ($GFT$) lowers routing overhead. While, OLSR due to Multi-Point Relay (MPRs) achieves more optimization in high densities by reducing the number of (re)transmissions. Each strategy implemented by a proactive protocol enhances the robustness of the protocol in any of the aspect(s) during different rates of mobility and node density.

	In this paper, different operations regarding link maintenance in proactive protocols are modeled and discussed. Moreover, different probabilities which affect the routing packet delivery fraction, are also modeled. In section II, related work and motivation are discussed. In section III and IV, a complete framework of modeling of routing overhead in proactive routing protocols is presented. Proposed framework is validated using NS-2 in section V.

%\vspace{-0.3cm}
\section{Related Work and Motivation}
For last few years, lot of studies have been done to cope with routing issues in WMhNs with the help of routing protocols. Proactive protocols, due to their periodic activities utilize more bandwidth, but in high densities, they perform well [4]. This is because of periodic computation of routes and local link/route maintenance with the neighbors result in low latency.

Tridib Mukherjee \textit{et al.} [6] aim to minimize energy wastage in the wireless network due to high \textit{control traffic}, which restricts the scalability and applicability of such protocols, without trading-off low latency. They present a model for optimum period for link and route maintenance as $\beta \_opt$ and $\varphi \_opt$, respectively.

Regarding routing overhead, Lin, T. \textit{et. al} [7] present an analytical model for comparison of routing protocols using overhead as a metric. They also apply their framework to improve a routing protocol by comparing the use of relay nodes of proactive protocols with flooding process of reactive protocols.

MPRs in [12] contain the concept of optimizing relaying node through collecting the degree and the connectivity of the second hop neighbors. Authors mention that how their strategy is fruitful for proactive protocols by evaluating OLSR with MPRs. They simulate OLSR with regular flooding and MPR flooding and deduce that MPRs are the key parameter for scalability of OLSR.

The work in [13] presents a model of key performance metrics of neighbor discovery algorithms, such as node degree and the distribution of the distance to symmetric neighbors. This model accounts for the dynamics of neighbor discovery as well as node density, mobility, radio and interference. Authors demonstrate a method for applying these models to the evaluation of global network metrics. In scrupulous, this paper describes a model of network connectivity.

Park \textit{et al.} in [14] give an expression; $C_{total}^{(rp)}=C_{E}^{(rp)} \times C_{T}^{(rp)}$ for total cost which is the product of the cost paid by a routing protocol for energy consumed per packet and time spent per packet. Authors mainly focus the comparison of simple flooding with Expanding Ring Search (ERS) algorithm. While, we model and evaluate $C_{E}^{(rp)}$ and $C_{T}^{(rp)}$ for DSDV, FSR and OLSR and compare the effects of their routing strategies on route table calculation. DSDV uses simple flooding, FSR uses multi-scope routing (MSR) (no flooding) while OLSR uses MRR flooding to calculate path. We define equations for measurement of different routing schemes for chosen protocols. Then we evaluate and validate them in different network scenarios using NS-2.

In [8], authors make survey about routing overhead of routing protocols. They characterize reactive and proactive protocols as "hello protocols" and "flooding protocols". They conclude from the simulations that more control packets are needed for hello protocols in mobile scenarios as compared to flooding protocols. Jacquet, P. \textit{et. al} [8] in their survey analysis, only discuss energy cost for routing protocols. While, we model both energy and time costs for proactive protocols. Moreover, modeling for MPRs, MSR with $GFT$, etc, is yet to be done. In our work, we, therefore, model the optimization methods for flooding in DSDV, FSR and OLSR.

Saleem \textit{et. al} [9] improve their work in [10], by taking inspiration from Broch, J. \textit{et. al} [11] and present flooding cost of routing protocols. They propose a performance evaluation framework that can be used to model two key performance metrics of an ad-hoc routing algorithm, namely routing overhead and route optimality. They also evaluate derivatives of two metrics; total energy consumption and route discovery latency for DSDV, DSR, AODV-LL and Gossiping. But they only model DSDV among proactive protocols, while we model FSR and OLSR along with DSDV with remarkable details. (We have modeled AODV, DSR and DYMO in [5]). They consider flooding based route discovery impacts over routing protocols, but we model MPR flooding in OLSR, MSR (no flooding) with $GFT$ in FSR along with simple flooding of DSDV. Their work is mainly concerned with stochastic probabilities; channel error and collision error. Contrary to their considered probabilities; we model probabilities of neighbor discovery errors. The reason for selecting neighbor discovery errors for DSDV, FSR and OLSR is that in proactive protocols efficient and quick neighbor detection leads to correct route discovery. We aim to analyze that which protocols estimate accurate route and best paths in which scenarios.

Neighbor discovery is a critical component of proactive routing protocols in wireless ad-hoc networks [13]. Neighborhood estimates are corrupted by two types of errors, namely \textbf{TypeI error:} occurs when a node believes that it has a neighbor when in fact it is not able to communicate with this node. \textbf{TypeII error:} occurs when a node is unaware that it is able to communicate with a node. These errors can have a significant impact on connectivity; if two nodes are unaware that they are neighbors, the link between them will not be known to the rest of the network. These errors are:

%eq1
\begin{eqnarray}
 P(Type I):= 1-\frac{\int_{0}^{d_{max}}p(sym,d)p_{pkt.suc(d)}dd}{\int_{0}^{d_{max}}p(sym,d)dd}
\end{eqnarray}

%eq2
\begin{eqnarray}
 P(Type I):= 1-\frac{\int_{0}^{d_{max}}p(sym,d)p_{pkt.suc(d)}dd}{\int_{0}^{d_{max}}p_{pkt.suc(d)}p(d)dd}
\end{eqnarray}

Referring the effects of errors on MAC layer in [13], we evaluate and compare the performance of routing protocols at network layer.

%\vspace{-0.3cm}
\section{Maintenance Operations of Proactive Routing Protocols}
There are three different \textit{operations} for maintaining network topology and route information in the proactive routing protocols; two are periodic and the third one is triggered.
%\vspace{-0.3cm}
\subsection{Maintenance Operations}
\textbf{a. Link Status Monitoring Periodically $(LSM\_Per)$:}
To maintain recent information about \textit{link status (LS)} in the network, a node needs to exchange information of establishment of links with its neighbors through periodically exchanging $LSM\_Per$. If a node does not receive any beacon message from a neighbor for a certain number of successive beacon periods, the link is assumed to be broken. Then routes are updated depending on the topology maintained by $LSM\_Per$.

\textbf{b. Route Updates Triggered for every change in the LS $(RU\_Tri)$:}
This operation updates routing information across the network whenever LS changes in an active route. Flooding of $RU\_Tri$ takes place to diffuse the updates across the network.
In rest of the paper, we use the terms 'broadcasting' and 'flooding' interchangeably.

\textbf{c. Updating Routes Periodically $(RU\_Per)$:}
Unlike $RU\_Tri$, this operation accumulates all link changes in a specified interval before broadcasting route updates.
We classify the proactive protocols in the following subsection based on the implementation of above mentioned operations in the protocols.

%\vspace{-0.3cm}
\subsection{Proactive Protocols with Basic Operations}
\textbf{a. DSDV:}
All of the aforementioned three operations are performed by DSDV. Although the $RU\_Tri$ operation may appear redundant because of the employment of $LSM\_Per$, it has certain consequences. $LSM\_Per$ in this protocol may lead to routing loops, which is corrected in $RU\_Per$ operations. $RU\_Per$ operation includes transmission of destination sequence numbers to monitor and maintain the freshness of the routing structures. After performing $LSM\_Per$, routing loops are removed by $RU\_Per$ with the latest sequence numbers.

\textbf{b. FSR:}
A moderate approach is taken in FSR, where $RU\_Tri$ is not performed at all. A main drawback of using both $LSM\_Per$ and $RU\_Tri$ is the large amount of control traffic generation. As, $RU\_Tri$ is performed with every change in the link status, so, it generates higher routing messages, especially during the high rates of mobility.

\textbf{c. OLSR:}
One of the main challenges using $RU\_Per$ with $LSM\_Per$ is to address the trade-off between amount of control traffic and the consistency of route information. OLSR performs only $RU\_Tri$ for maintaining fresh routes. Unlike DSDV, this protocol does not rely on destination sequence numbers in maintaining fresh loop-free routes.

%\vspace{-0.3cm}
\section{Modeling Routing Overhead Generated by Proactive Protocol}
In this section, we are presenting the detailed modeled equations for route maintenance operations for the selected proactive protocols. These equations mainly concern with routing overhead in terms of energy consumed and time spent per packet. A routing protocol $(rp)$ has to pay some cost in the form of consumed energy (for routing packet) for route calculation; $C_E^{(rp)}$, and $C_T^{(rp)}$ is time cost for computing end-to-end path.

In [14], authors have expressed the total cost paid by routing protocols; $C_{total}^{(rp)}$ by the following equation:

%eq3
\begin{eqnarray}
C_{total}^{(rp)}=C_{E}^{(rp)} \times C_{T}^{(rp)}
\end{eqnarray}

In following subsections, we model the routing overhead generated by three proactive protocols.
%\vspace{-0.3cm}
\subsection{Modeling Overhead by DSDV}
In DSDV, $RU\_Tri$ are generated if a link breaks among active route. Moreover, after a specific time period, route table updates are also periodically advertised; $C_{E-Per}^{(DSDV)}$. A single network protocol data unit (NPDU) is enough to spread the routing information in the network when the network is neither dense nor dynamic. On the other hand, in case of high dynamic environment, route table information is advertised in multiple NPDUs per $RU\_Per$. Flooding process is used to spread the information in DSDV. NPDUs are used to broadcast the distance vector information. Total energy cost for DSDV; $C_{E-total}^{(DSDV)}$ is the sum of energy consumed for spreading periodic updates $(LMS\_Per$ and $RU\_Per)$; $C_{E-Per}^{(DSDV)}$ and energy cost to spread $RU\_Tri$; $C_{E-Tri}^{(DSDV)}$ (eq.4).

%eq4
\begin{eqnarray}
C_{E-total}^{(DSDV)}=C_{E-Per}^{(DSDV)} + C_{E-Tri}^{(DSDV)}
\end{eqnarray}

%eq5
\begin{eqnarray}
C_{E-Per}^{(DSDV)}=\int _{0}^{\tau _{LPU}}(P_{err}d_{avg} + d_{avg} \sum _{i=0}^{h-1}(P_{err})^{i+1}\prod _{j=1}^{i}d_f[j])
\end{eqnarray}

%eq6
\tiny
\begin{eqnarray}
C_{E-Tri}^{(DSDV)}=\int _{0}^{\tau_{LTU}}\sum_{p=1}^{M} \sum_{n=1}^{N} (1-P_{nlb})_n P_{err}d_{avg} + d_{avg} \sum _{i=0}^{h-1}(P_{err})^{i+1}\prod _{j=1}^{i}d_f[j]
\end{eqnarray}
\normalsize

We define two probabilities for broadcasting trigger and periodic updates; (i) $P_{err}$: which might be $P(Type I)$ in eq.1 or $P(Type II)$ in eq.2 [15], (ii) $p_{nlb}$: which is no link breakage probability or nodes' stability. $d_{avg}$ and $d_f$ indicate average and forwarding degree of a node in the network, while $h$ is the maximum Time-To-Live $(TTL)$ value of broadcasting. Readers are advised to consult chapter 3 in [4] to better understand eq.5. $\tau_{LPU}$ and $\tau_{LTU}$ specify time periods for last periodic updates and last trigger updates. $p$ in the $M^{th}$ path and $n$ is the $N^{th}$. $C_T^{(DSDV)}$ is defined below:

%eq7
\small
\begin{eqnarray}
C_T^{(DSDV)}=
 \begin{cases}
   \tau_{rst}+\displaystyle\sum_{i=1}^{node} L & successful\,packet\, delivery \\
   \tau_{rst}+\displaystyle\sum_{i=1}^{node_{lb}} L+\tau_{TU}       & otherwise
  \end{cases}
\end{eqnarray}
\normalsize

Where, $node$ presents a node in an active path, $node_{lb}$ is the node which detects a link break $(lb)$, and $L$ is the transmission delay over a single link. $L$ becomes important during the states of congestion, interference, etc. If the time of delivery is important, carried information can be out-of-date at the moment it is received and used by a node [15]. Unlike other proactive protocols, DSDV keeps a data for duration of \textit{route settling time} $(\tau_{rst})$ to provide a stabilized route. This strategy augments delay but it guarantees an accurate path, especially in case of high dynamic environment. For a $\tau _{rst}$, $\tau _{tu}$ is the respective \textit{trigger update} time for providing the correct routes in routing tables.

%\vspace{-0.3cm}
\subsection{Modeling Overhead by FSR}
In FSR, flooding is not used but broadcasting in inner-scope and outer-scope is performed through MSR mechanism to reduce the routing overhead. TTL value for broadcasting is set to respective periodic scope, so, flooding does not take place. In routing, fish-eye approach using MSR with $GFT$ maintains accurate distance and path quality with progressively less detail as the distance increases.

When network size grows, the update message could consume considerable amount of bandwidth, which depends on the update period. $C_E^{(FSR)}$ is defined above, which is the sum of energy consumed per route packet generation in inner-scope plus outer-scope. Energy cost for inner scope; $C_E^{in-sco}$ and for outer scope; $C_E^{out-sco}$, are given in eq.9, 10.

In eq.8, $\tau_{LSU}$ is the \textit{last scope update} period, $N_{out}$ and $N_{in}$ are the TTL values for outer and inner scopes, respectively. In FSR, in case of link failure, no trigger event takes place, routing table is updated on the \textit{next scope update period}; $\tau _{nsu}$. So, $C_T^{(FSR)}$ can be written as:

%eq8

\begin{eqnarray}
C_E^{(FSR)}=\int _{0}^{\tau_{LSU}}C_E^{in-sco}+C_E^{out-sco}
\end{eqnarray}

%eq9
\begin{eqnarray}
C_E^{out-sco}=d_{avg}^{out}\sum_{i=1}^{N_{out}-1} (p_{err})^{i+1}\prod _{j=1}^{i}d_f[j]
\end{eqnarray}

%eq10
\begin{eqnarray}
C_E^{in-sco}=d_{avg}^{in}\sum_{i=1}^{N_{in}-1} (p_{err})^{i+1}\prod _{j=1}^{i}d_f[j]
\end{eqnarray}

%eq11
\begin{eqnarray}
C_T^{(FSR)}=
 \begin{cases}
   \displaystyle\sum_{i=1}^{node} L & successful\,packet\, delivery \\
   \tau_{nsu}       & otherwise
  \end{cases}
\end{eqnarray}

%\vspace{-0.3cm}
\subsection{Modeling Overhead by OLSR}
In this protocol, two types of periodic control messages are used: Topology Control (TC) and HELLO messages. The former is used for getting the whole topology map, while the later is used for exchanging symmetric information of the neighbors and calculating MPRs. So,

%eq12
\begin{eqnarray}
C_{E}^{(OLSR)}=C_{E-TC}^{(OLSR)} + C_{E-HELLO}^{(OLSR)}
\end{eqnarray}

After every TC interval, the status of $MPRs$ is checked that mainly depends on stability of nodes. $p_c^{MPR}$ denotes the probability of change in MPRs. Flooding of next TC message depends upon this probability.

%eq13
\begin{eqnarray}
C_{E-TC}^{(OLSR)}=\int _{0}^{\tau_{LU}^{TC}} C_{E-nc}^{MPR} + C_{E-c}^{MPR}
\end{eqnarray}

Where, $\tau_{LU}^{TC}$ is the $last update$ of generation of TC messages. $C_{E-nc}^{MPR}$ is the cost of allowed (re)transmissions through MPRs, while $C_{E-c}^{MPR}$ shows the cost of dissemination of TC messages in the whole network.

%eq14
\tiny
\begin{eqnarray}
C_{E-nc}^{MPR}=(1-p_c^{MPR})p_{err}d_{avg}+d_{avg}\sum_{i=1}^{h-1} (p_{err})^{i+1}\prod _{j=1}^{i}d_f^{MPR}[j]
\end{eqnarray}
\normalsize

%eq15
\small
\begin{eqnarray}
C_{E-c}^{MPR}=p_c^{MPR} p_{err}d_{avg}+d_{avg}\sum_{i=1}^{h-1} (p_{err})^{i+1}\prod _{j=1}^{i}d_f[j]
\end{eqnarray}
\normalsize

%eq16
\begin{eqnarray}
C_{E-HELLO}^{(OLSR)}=\int _{n=0}^{N} (p_{err})_n d_{nb}[n]
\end{eqnarray}

$d_{nb}[n]$ is the degree of neighbor nodes and $N$ is the total number of nodes in the network. $d_f^{MPR}$ is the MPRs forwarding degree. Like [12], we define $d_f^{MPR}$: if $x$ represents a node in the network, and $N(x)$ is the set of first-hop neighbor and $N^2(x)$ are the  second hop neighbors of node $x$, then $N(x)$ is in the neighborhood of $x$, i.e., $x\notin N(x)$. If $y$ is the first-hop neighbor of $x$, it means $x$ covers $y$ or merely $y$ is the neighbor of $x$. Further, if $S$ and $T$ are set of nodes, then $x$ covers $T$ $iff$ every node of $T$ is covered by some node present in $S$. A set of first-hop neighbor $S\subseteq N(x)$ is $MPR$ of node $x$ if it wraps $N^2(x)$, or equivalently $\cup _{y\in N(x)}N(y)-N(x)\subseteq \cup_{y\in S} N(y)$. So, MPR forwarding degree can be calculated as:
\newline
$\cup _{y\in N(x)}N(y)-N(x)\subseteq \cup_{y\in S} N(y)$
\newline

%eq17
\small
\begin{eqnarray}
C_T^{(OLSR)}=
 \begin{cases}
   \displaystyle\sum_{i=1}^{node} L & successful\,packet\, delivery \\
   \displaystyle\sum_{i=1}^{node} L+\tau_{dlb}+\tau_{Trig}^{TC}      & otherwise
  \end{cases}
\end{eqnarray}
\normalsize

Where, $\tau_{dlb}$ is the time period to \textit{detect the link breakage} after consecutive failure of link sensing. $\tau_ {Trig}^{TC} $ may be equal or less than $RU-Tri$ period, as, it depends on $p_c^{MPR}$.

%\vspace{-0.3cm}
\section{Simulations}
We consider three performance parameters; RREQ Packet Generation, average E2ED, and average NRL. Both common and different parameters are given in table.1 and in table.2. The simulation results are shown in fig.1 and are summarized in table.3. E2ED and NRL reductions have already been achieved with two new routing link metrics; Interference and bandwidth adjusted ETX (IBETX) [16] and Inverse ETX [17].

%Table.1. Common Simulation Parameters
    %\vspace{-0.5cm}
      \begin{table}[!h]
    \centering
    \small
    \begin{tabular}{|c|c|}
    \multicolumn{2}{c}{Table.1. Common Simulation Parameters} \\
    \hline
    \textbf{Parameter}&\textbf{Value}\\
    \hline
    Area & 1000 x 1000 $m^2$\\ \hline
    Simulation Time& 900 Seconds \\ \hline
    Data Traffic Source & CBR of 512 bytes \\
    \hline
    Mobility Model&Random Way Point\\
    \hline
    Wireless Link Bandwidth&2Mbps\\
    \hline
    Speed&15mps\\
    \hline
    \end{tabular}
    \normalsize
     \end{table}

 %Table.2. Different Simulation Parameters
     \begin{table}[!h]
    \centering
    \small
    \begin{tabular}{|c|c|c|}
    \multicolumn{3}{c}{\hspace{-3cm} Table.2. Different Simulation Parameters} \\
    \hline
    \textbf{Scenario}& \textbf{No. of Nodes}&\textbf{Pause Time (s)}\\ \hline
    Mobilities (Sc.1)& 100 &0,100,200,300\\
    \hline
    Scalabilities (Sc.1) & 100,150,200,250&2 \\
    \hline
    Hi Mobilities (Sc.2)&50&0,100,200,300,400\\
    \hline
    Lo Mobilities (Sc.2)&50&500,600,700,800,900\\
    \hline
    Lo Scalabilities (Sc.3)&10,20,30,40,50&2\\
    \hline
    Hi Scalabilities (Sc.3)&60,70,80,90,100&2\\
    \hline
    \end{tabular}
    \normalsize
     \end{table}

%\begin{sidewaystable}[!h]
    \begin{table*}[!t]
    \centering
    \resizebox{16cm}{4cm}{
    %\tiny
    \begin{tabular}{|c|c|p{4cm}|p{6cm}|p{2cm}|}
    %\hline
    \multicolumn{5}{c}{\hspace{-14cm} Table.3. Simulation Results} \\
    \hline
    \textbf{Parameters}&\textbf{Protocols}&\textbf{Findings}&\textbf{Reasons}&\textbf{Ref. of Eq.}\\
    \hline
    \multicolumn{1}{|c|}{\multirow{3}{*}{\textbf{Routing Overhead}}} &
\multicolumn{1}{|c|}{\textbf{DSDV}}& Fig.1.a,b. Increase in high scalabilities and high mobilities & High value of $p_{nlb}$ & eq.6        \\ \cline{2-5}
\multicolumn{1}{|c|}{}                        &
\multicolumn{1}{|c|}{\textbf{FSR}} & Lowest number of routing packets, depicted in Fig.1.a,b & Generation of trigger updates both in DSDV and OLSR produces more number of control packets in varying mobility and scalability as compared to FSR & eq.5,6,15 compared to eq.10        \\ \cline{2-5}
\multicolumn{1}{|c|}{}                        &
\multicolumn{1}{|c|}{\textbf{OLSR}} & Producing highest routing packet overhead (Fig.1.a,b) & $HELLO$ and $TC$ message generation during $RU\_Tri$ & eq.13 and 16        \\ \cline{1-5}

\multicolumn{1}{|c|}{\multirow{3}{*}{\textbf{E2ED}}} &
\multicolumn{1}{|c|}{\textbf{DSDV}}& The highest E2ED in Fig.1.c,d & (1) It keeps a data packet until it receives a good route, i.e. for $\tau _{rst}$ (2) Advertisement of the routes which are not stabilized yet, is delayed in order to reduce number of rebroadcasts of possible route entries
 & eq.7        \\ \cline{2-5}
\multicolumn{1}{|c|}{}                        &
\multicolumn{1}{|c|}{\textbf{FSR}} & The highest AE2ED in Fig.1.c & (1) Due to GFT, probabilities; $p^{in-sco}$ and $p^{out-sco}$ (2) Due to absence of $RU\_Tri$, or Lack of any mechanism to delete expired stale routes or to determine freshness of routes when multiple routes are available in $RC$
 & eq.9,10        \\ \cline{2-5}
\multicolumn{1}{|c|}{}                        &
\multicolumn{1}{|c|}{\textbf{OLSR}} & Overall lowest E2ED in Fig.1.c,d,g,h  & Generates $periodic HELLO$ and $triggered TC$ messages to check links to compute the MPRs to better reduce delay as compared to $RU\_Per$ of whole table with neighbors in FSR and $RU\_Per$ and $RU\_Tri$ in DSDV & eq.14,15,16 compared to eq.5        \\ \cline{1-5}

\multicolumn{1}{|c|}{\multirow{3}{*}{\textbf{NRL}}} &
\multicolumn{1}{|c|}{\textbf{DSDV}}& Lowest NRL in Fig.1.e,f and Fig.1.i. Increasing NRL in Fig.1.j & $RU\_Tri$ and $RU\_Per$ through NPDUs reduce routing overhead. In small population, chance of full dump is reduced. So, NPDUs produce routing packets that gradually increase due to small population of nodes to the large one.& eq.5,6. \\ \cline{2-5}
\multicolumn{1}{|c|}{}                        &
\multicolumn{1}{|c|}{\textbf{FSR}} & Maintains lower routing overhead than OLSR Fig.1.e,f and 1.i. Lowest NRL shown in Fig.1.j
 &GFT mechanism with MSR & eq.9,10        \\ \cline{2-5}
\multicolumn{1}{|c|}{}                        &
\multicolumn{1}{|c|}{\textbf{OLSR}} & Highest NRL in Fig.1.e,f and Fig.1.i,j & Computation of MPRs through $TC$ and $HELLO$ messages results in the highest generation of routing packets (trade-off between NRL and E2ED) & eq.13,14,15,16        \\ \cline{1-5}
    \end{tabular}
    }
    \end{table*}
    %\normalsize
    %\end{sidewaystable}

    %\vspace{-0.3cm}
\section{Conclusion and Future Work}

In this work, we have considered three wireless proactive routing protocols; DSDV, FSR and OLSR. We further classify these protocols on the basis of their routing strategies. As wireless nodes running proactive protocols periodically exchange local information with their neighbors, and build a complete topological map on the basis of this information, a most important factor is the probability of accurate measurement of neighbor detection. So, in our framework, we calculated the probabilities of changed MPRs, neighbor,discovery error and no link breakage. The situations during which NPDUs are generated in the case of DSDV, MPR redundancy is applied by OLSR and route request broadcasts are limited through MSR using $GFT$ in FSR, are modeled in our proposed equations. In future, we are interested to reduce E2ED and routing load by implementing new link metrics with FSR, like IBETX and InverseETX are implemented with DSDV [16] and OLSR [17], respectively.

%\vspace{1cm}

    \begin{figure}[!h]
  \centering
 \subfigure[Mobilities in Scenario.1]{\includegraphics[height=2 cm,width=4.3 cm]{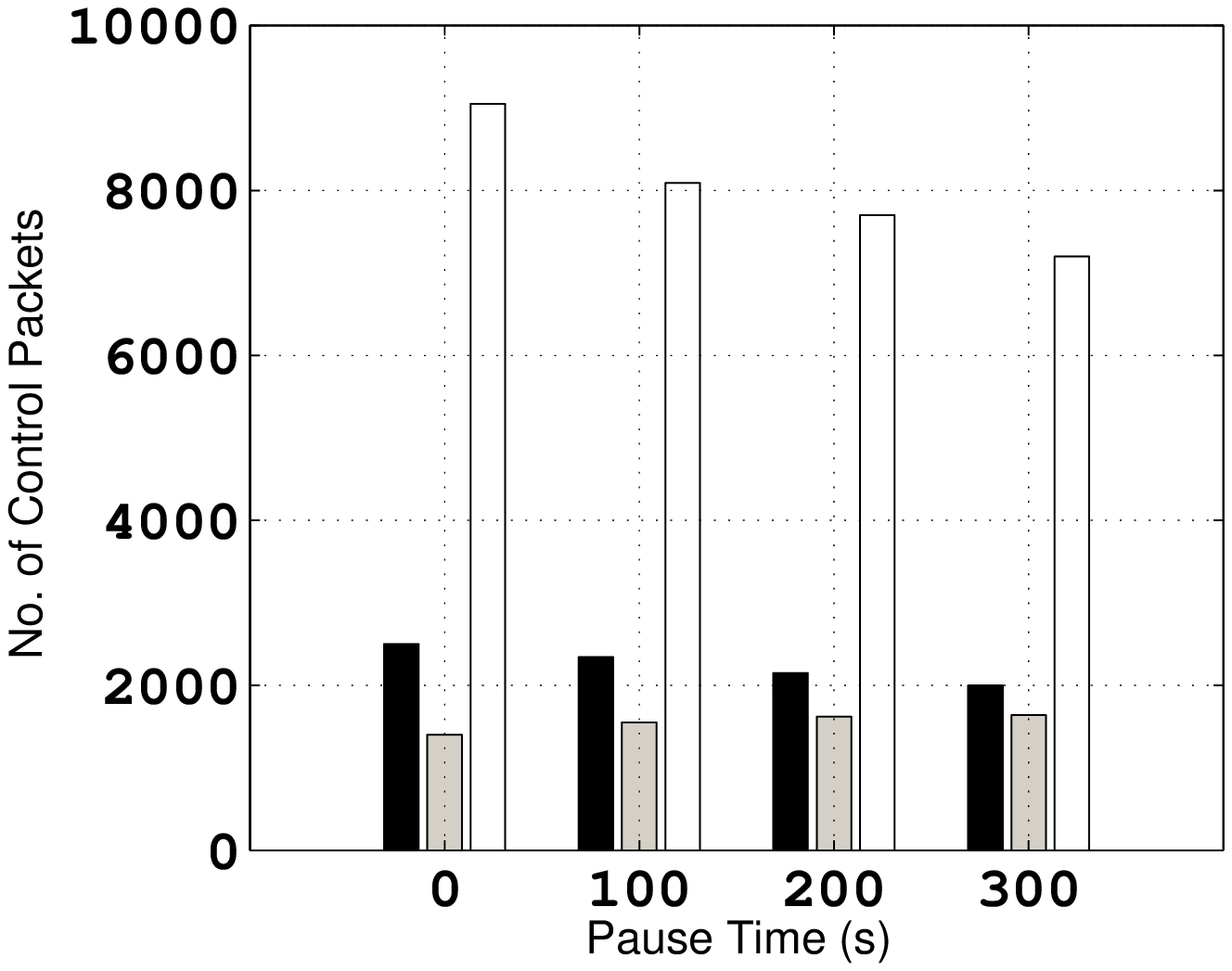}}
 \subfigure[Scalabilities in Scenario.1]{\includegraphics[height=2  cm,width=4.3 cm]{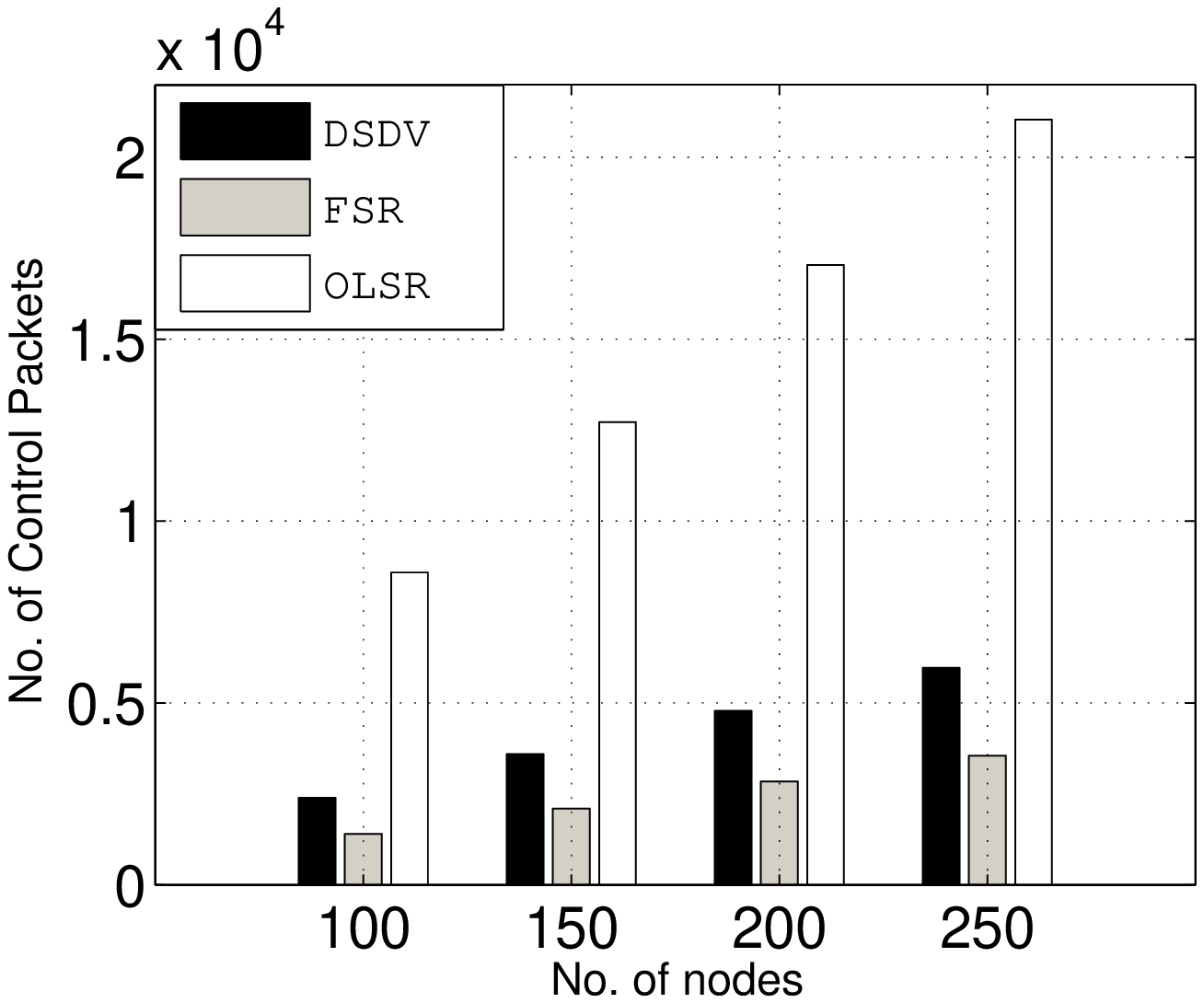}}
 \subfigure[Hi mobilities in Scenario.2]{\includegraphics[height=2 cm,width=4.3 cm]{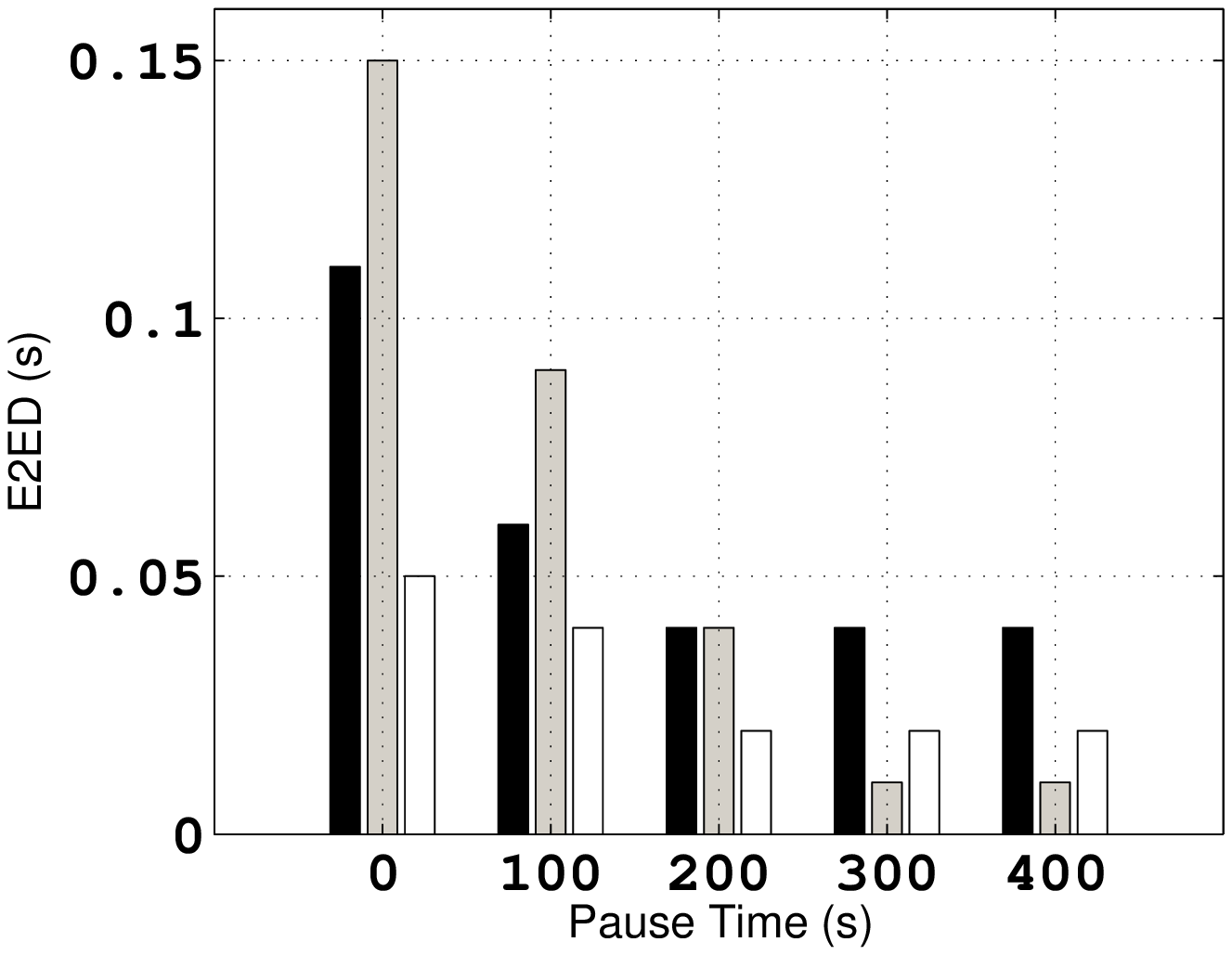}}
 \subfigure[Lo mobilities in Scenario.2]{\includegraphics[height=2  cm,width=4.3 cm]{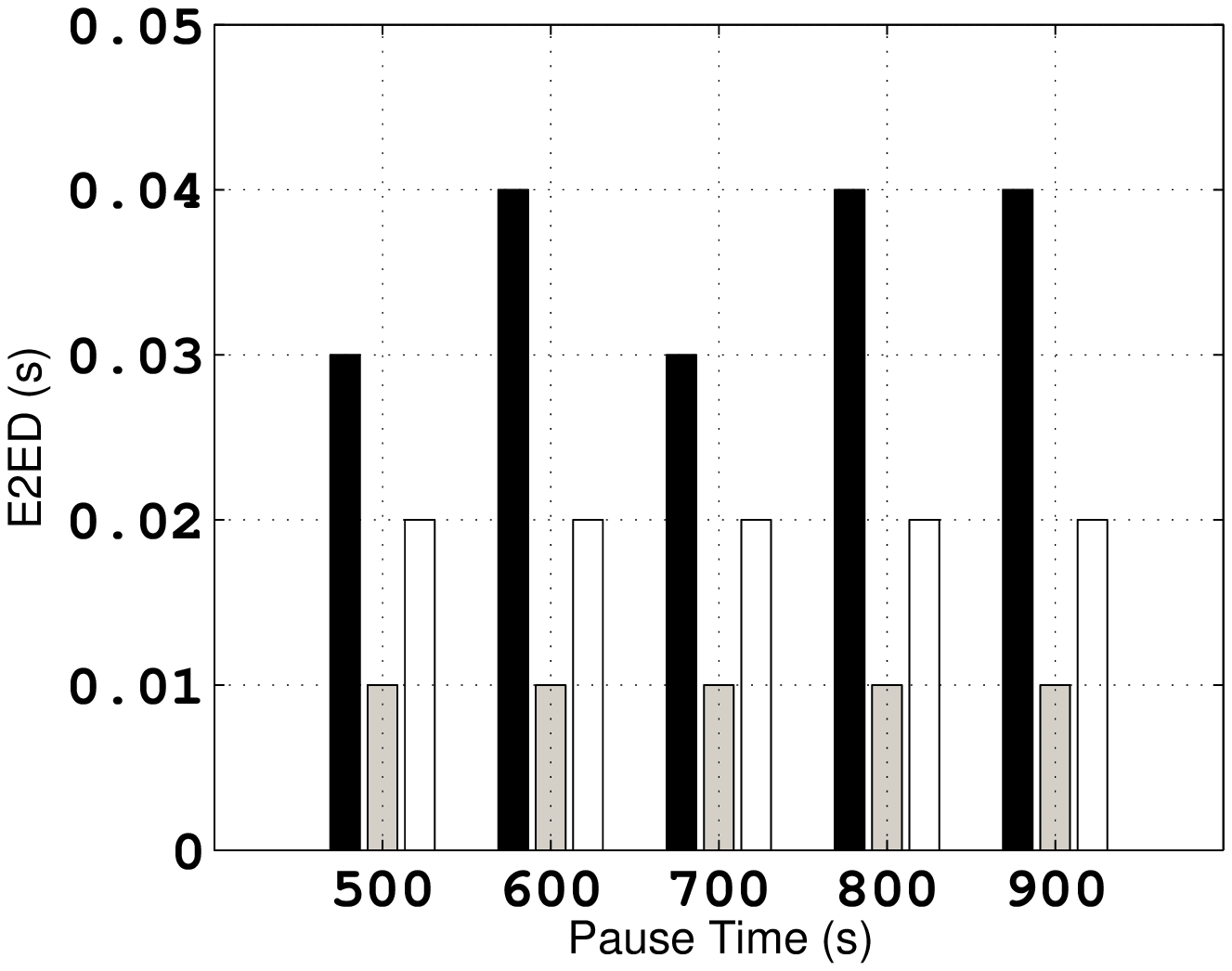}}
  \subfigure[Hi mobilities in Scenario.2]{\includegraphics[height=2 cm,width=4.3 cm]{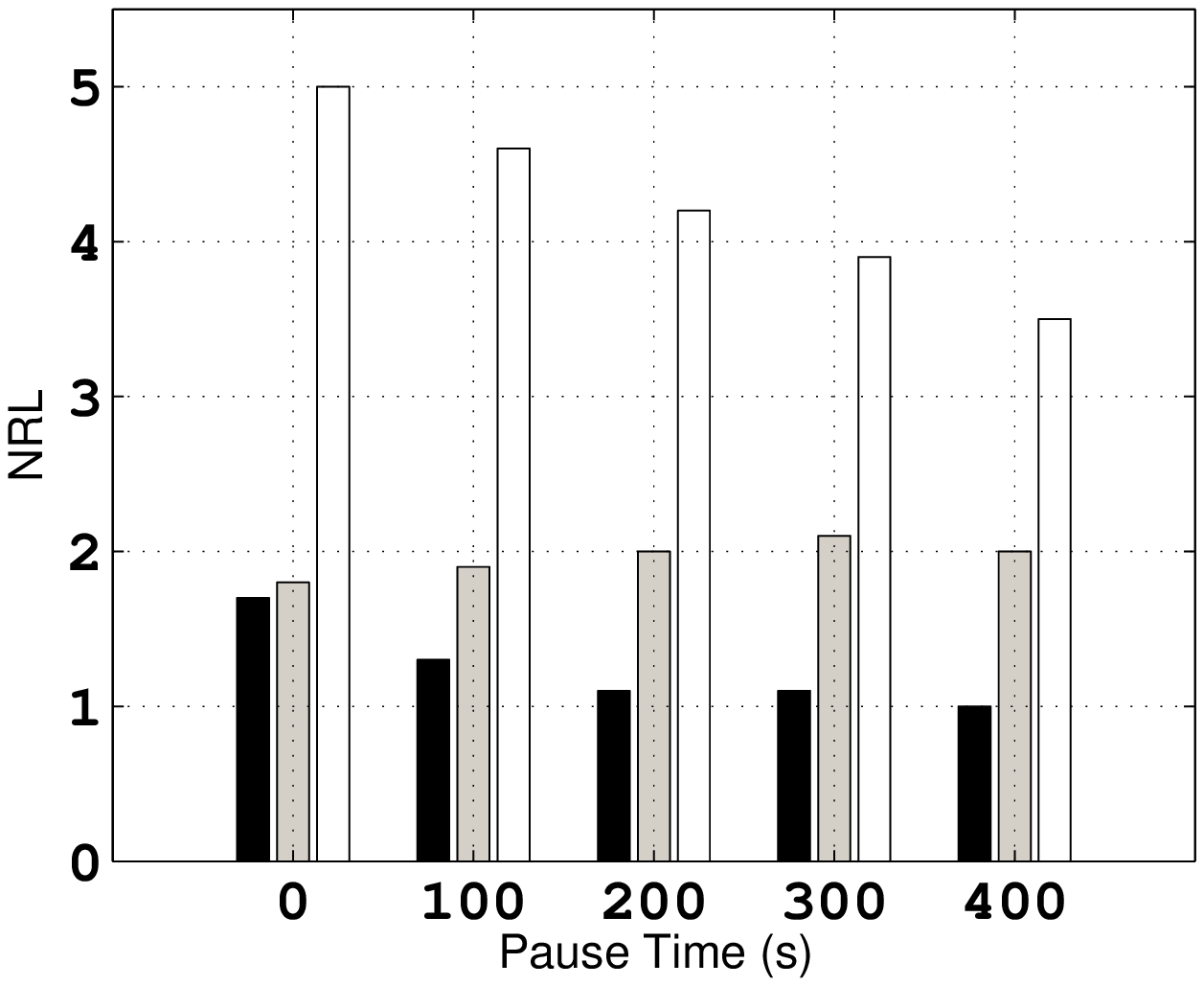}}
 \subfigure[Lo mobilities in Scenario.2]{\includegraphics[height=2  cm,width=4.3 cm]{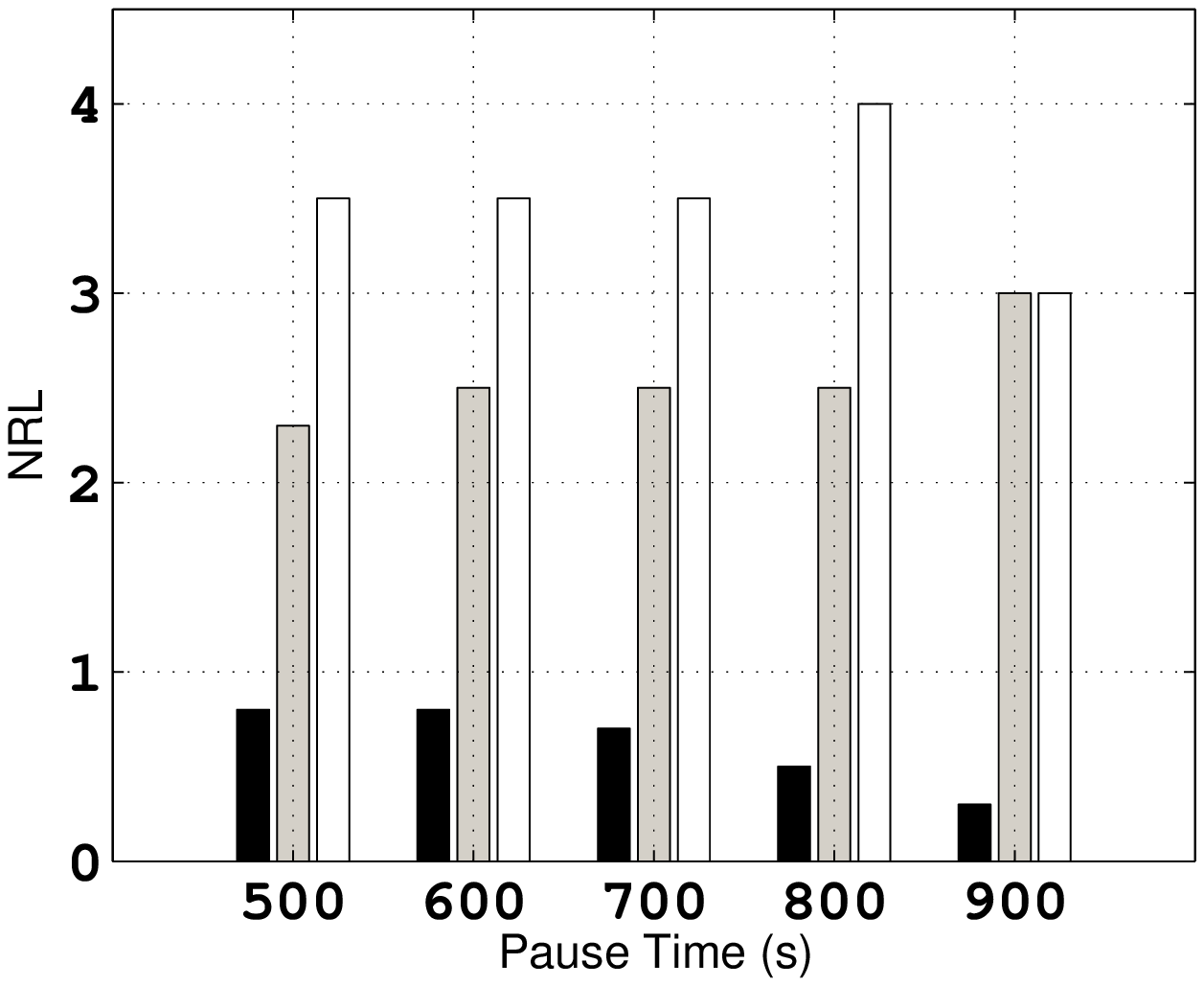}}
 \subfigure[Lo scalabilities in Scenario.3]{\includegraphics[height=2 cm,width=4.3 cm]{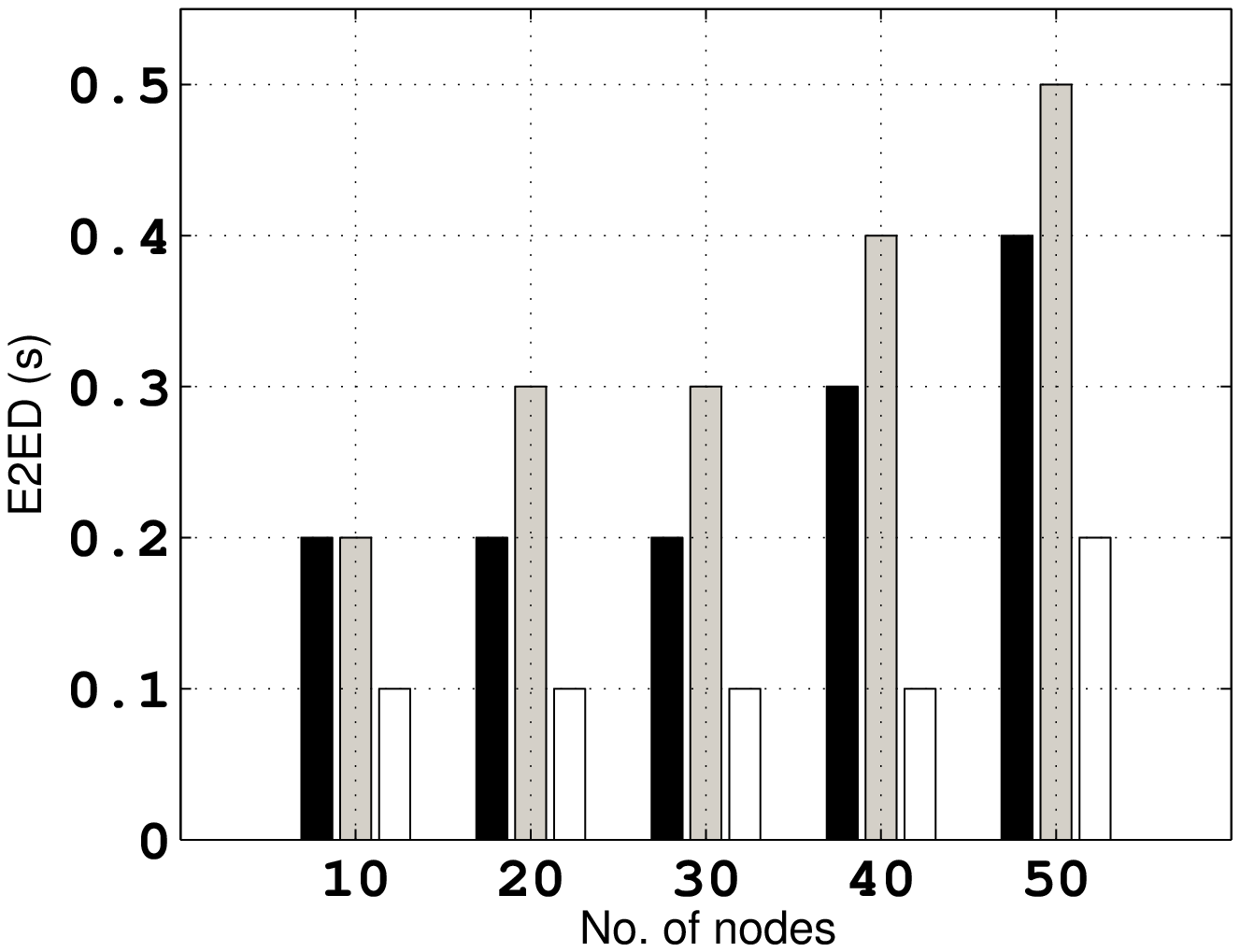}}
 \subfigure[Hi scalabilities in Scenario.3]{\includegraphics[height=2  cm,width=4.3 cm]{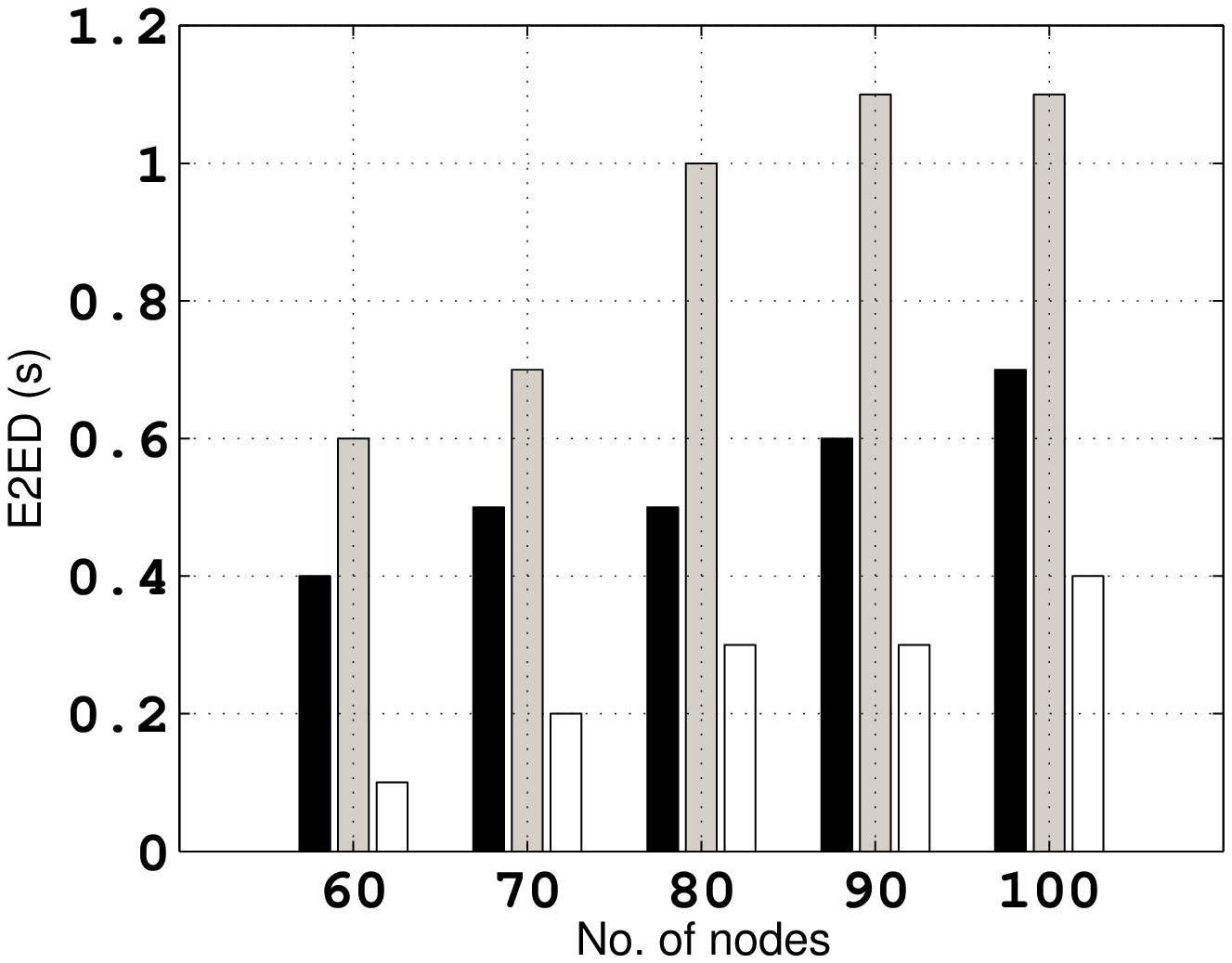}}
  \subfigure[Lo scalabilities in Scenario.3]{\includegraphics[height=2 cm,width=4.3 cm]{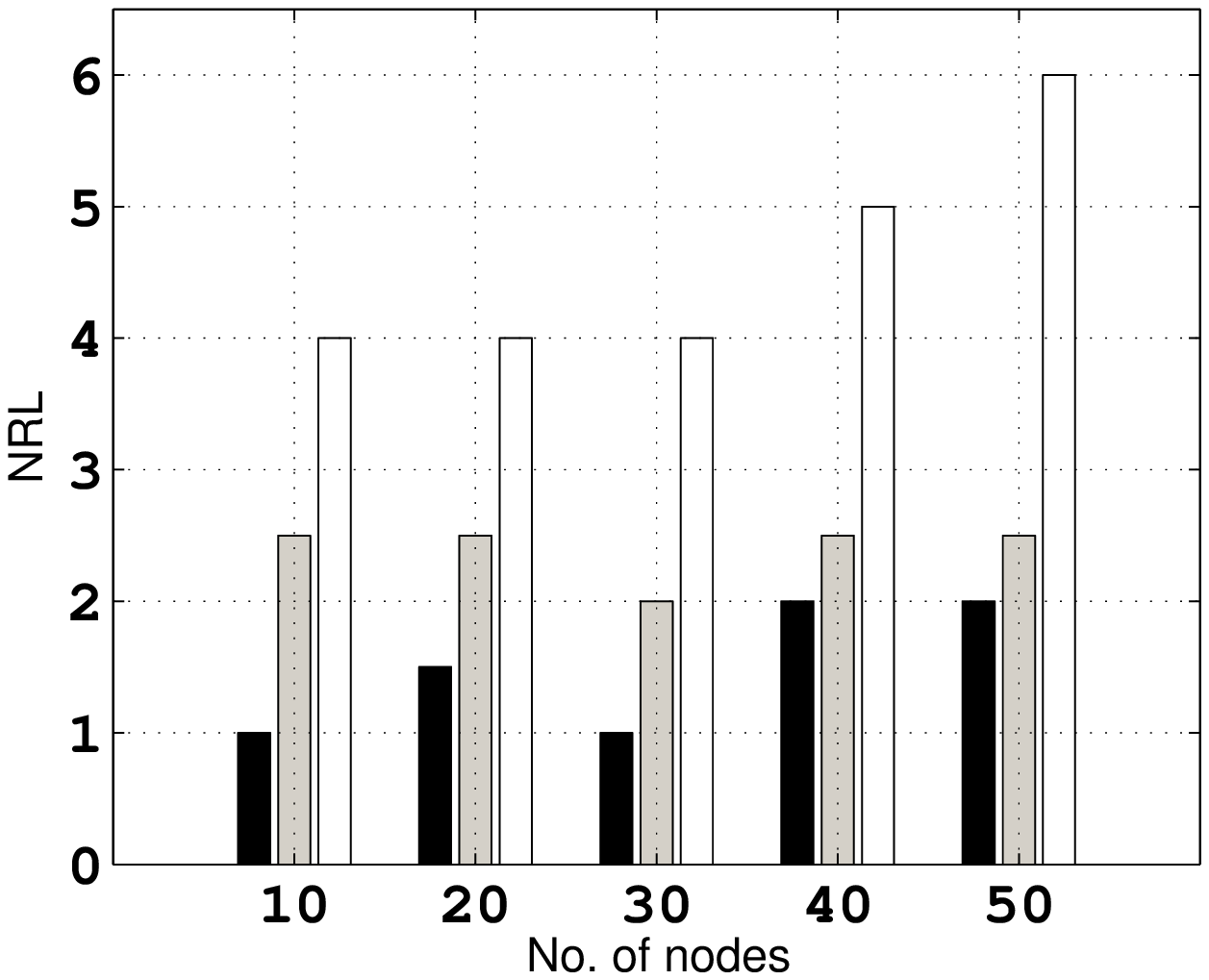}}
 \subfigure[Hi scalabilities in Scenario.3]{\includegraphics[height=2  cm,width=4.3 cm]{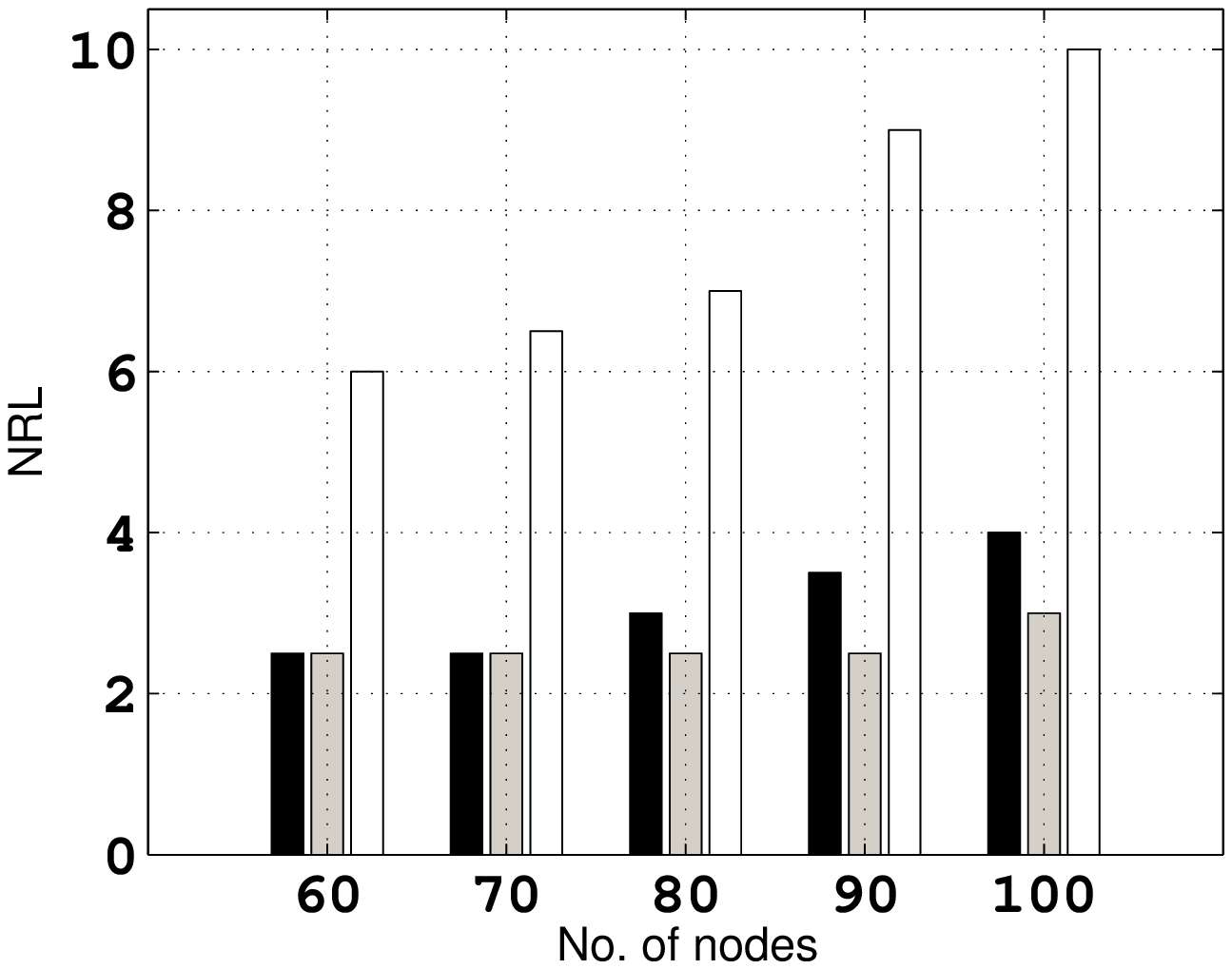}}
  \caption{Simulation Results for Modeled Framework}
\end{figure}

\end{document}